\documentclass[aps,twocolumn,superscriptaddress,raggedbottom,showpacs,floatfix]{revtex4-1}
\usepackage[pdftex]{graphicx}
\usepackage{amssymb}
\usepackage{amsmath}
\usepackage{color}
\usepackage{hyperref}
\usepackage{textcomp}
\usepackage{booktabs}
\hypersetup{colorlinks=true,linkcolor=blue,urlcolor=blue,citecolor=blue}

\newcommand{\cT}{{\cal T}}
\newcommand{\cP}{{\cal P}}

\begin{document}

\title{Giant tunable nonreciprocity of light in Weyl semimetals}
\author{O.~V. Kotov}%
\email{oleg.v.kotov@yandex.ru}%
\affiliation{N.~L. Dukhov Research Institute of Automatics (VNIIA), 127055 Moscow, Russia}%
\author{Yu.~E. Lozovik}% 
\email{lozovik@isan.troitsk.ru}%
\affiliation{Institute for Spectroscopy, Russian Academy of Sciences, 142190 Troitsk, Moscow, Russia}%
\affiliation{N.~L. Dukhov Research Institute of Automatics (VNIIA), 127055 Moscow, Russia}%
\affiliation{National Research University Higher School of Economics, 101000 Moscow, Russia}

\begin{abstract}
The propagation of light in Weyl semimetal films is analyzed. The magnetic family of these materials is known by anomalous Hall effect, which, being enhanced by the large Berry curvature, allows one to create strong gyrotropic and nonreciprocity effects without external magnetic field. The existence of nonreciprocal waveguide electromagnetic modes in ferromagnetic Weyl semimetal films in the Voigt configuration is predicted. Thanks to the strong dielectric response caused by the gapless Weyl spectrum and the large Berry curvature, ferromagnetic Weyl semimetals combine the best waveguide properties of magnetic dielectrics or semiconductors with strong anomalous Hall effect in ferromagnets. The magnitude of the nonreciprocity depends both on the internal Weyl semimetal properties, the separation of Weyl nodes, and the external factor, the optical contrast between the media surrounding the film. By tuning the Fermi level in Weyl semimetals, one can vary the operation frequencies of the waveguide modes in THz and mid-IR ranges. Our findings pave the way to the design of compact, tunable, and effective nonreciprocal optical elements. 
\end{abstract}

\maketitle
\section{Introduction} \label{Sec1}
Weyl semimetals (WSs), being topologically nontrivial phase of matter, have recently attracted significant attention due to their massless bulk fermions and protected Fermi arc surface states with the corresponding topological
transport phenomena \cite{Volovik,Murakami,Halasz-Balents,Burkov-Balents,HosurQi:rev}. WS band structure
contains an even number \cite{Nielsen-Ninomiya} of nondegenerate band-touching points (Weyl nodes), which are topologically stable and can be regarded as magnetic monopoles and antimonopoles in the momentum space with positive or negative chiral charges and corresponding nonzero Chern numbers acting as the source and drain for the Berry curvature field \cite{Hasan, QiHughes}. The topological protection of massless fermions in WSs against weak perturbations follows from the separation of the individual Weyl nodes with opposite topological charges in momentum space, as the chiral Weyl nodes can only be destroyed by chirality mixing, which requires two opposite chirality Weyl nodes to meet. Such a separation demands breaking of either time-reversal ($\cT$) or inversion ($\cP$) symmetry, or both \cite{Murakami}. In WSs with lack of $\cP$ symmetry, the Weyl nodes separation is roughly proportional to the strength of the spin-–orbit coupling (SOC), which indicates the crucial role played by SOC in the formation of WSs \cite{Liu2015}. By contrast, in $\cT$- and $\cP$-invariant bulk Dirac semimetals (BDSs) where, according to Kramers theorem, all bands are doubly degenerate, the massless bulk fermions require additional crystal symmetries to be stable \cite{YangNagaosa}. 

The realization of a BDS phase in $\textrm{Na}_3\textrm{Bi}$, $\textrm{Cd}_3\textrm{As}_2$, and $\textrm{ZrTe}_5$ compounds was predicted \cite{Wang2012,Wang2013} and confirmed experimentally \cite{Liu:Science,Borisenko,Neupane,Liu:Nature,QLi}.
WS phase natural realizations contain the family of $\cT$-broken magnetic materials including pyrochlore iridates $\textrm{Y}_2\textrm{IrO}_7$, $\textrm{Eu}_2\textrm{IrO}_7$ \cite{Wan,Sushkov_opt_exp}, ferromagnetic spinels $\textrm{HgCr}_2\textrm{Se}_4$ \cite{Xu_spinels}, and Heusler ferromagnets $\textrm{Co}_3\textrm{S}_2\textrm{Sn}_2$, $\textrm{Co}_3\textrm{S}_2\textrm{Se}_2$ \cite{WangLei,LiuFelser,XuFelser}. This family also includes spin gapless compensated ferrimagnets $\mathrm{Ti}_{2}\mathrm{MnAl}$, where, in contrast to ferromagnetic WSs, the spin degeneracy is broken even without SOC, and  $\cT$-broken WS phase exist despite a zero net magnetic moment \cite{Shi}. The WS family of $\cP$-broken nonmagnetic materials includes noncentrosymmetric compounds TaAs, TaP, NbAs, and NbP \cite{Huang_arcs,Weng_PRX,Shekhar,Lv_TaAs,Behrends,Xu_TaAs,Xu_NbAs,Xu_TaP,Xu_opt_exp} (the detailed WS classification can be found in Refs.~\cite{RevDirac2016,RevHasan2017,RevArmitage2017}). Moreover, in some compounds, e.g., $\textrm{WTe}_2$ \cite{Soluyanov_II,Wang_II} and $\textrm{MoTe}_2$ \cite{Sun_II,Tamai_II,Huang_II}, the tilt of the Weyl cones exceeds the Fermi velocity giving rise to a type--II WS with open Fermi surface and a different type of Weyl fermions at the boundary between electron and hole pockets \cite{Xu_II,Sun_II,Soluyanov_II}.  

The nontrivial bulk band topology of WSs manifests in a number of exotic physical effects such as the protected against weak perturbations Fermi arc surface states \cite{Lee,Sun,Ojanen,Hosur,Potter} that connect the projections of the Weyl nodes in the surface Brillouin zone, the chiral anomaly \cite{Adler,Bell-Jackiw,Nielsen-Ninomiya,Aji,Parameswaran} (nonconservation
of the chiral charge transferred between Weyl nodes of opposite chirality), and related negative longitudinal magnetoresistance \cite{HosurQi:rev,Huang_Anomaly,Zhang_anomaly} quadratic in magnetic field, which appears if parallel electric and magnetic fields are applied. Also, WSs possess two basic phenomena related to the chiral anomaly: the chiral magnetic effect (CME)  \cite{Burkov_CME,QLi,HosurQi:rev,MaPesin,Baireuther,ZhouChang} and the anomalous Hall effect (AHE) \cite{HosurQi:rev,Burkov_AHE,Yang_AHE,Zyuzin_AHE}, which are closely related to the topological magnetoelectric effect in $\cT$-invariant
topological insulators \cite{QiHughes}. The CME, manifested in $\cP$-broken WSs as the electrical currents induced along the magnetic field, hypothetically could be caused by only a magnetic external field and not be associated with the chiral anomaly \cite{ZyuzinBurkov}. However, in an equilibrium state, when all contributions from filled electronic states are taken into account, the static magnetic-field-driven current must vanish \cite{VazifehFranz}. Thus, the nonvanishing CME implies the nonzero chiral chemical potential (the difference between local chemical potentials in Weyl nodes), which can be realized only in the nonequilibrium state dynamically generated by DC parallel electric and magnetic fields and associated with the chiral anomaly \cite{MacDonald,RevArmitage2017}. While the dynamic CME with the violation of the chiral current conservation is the consequence of the chiral anomaly, the AHE in any $\cT$-broken system, being the Hall effect in the absence of a magnetic field, strictly speaking, may be not a part of the chiral anomaly in WSs. We underline that due to the nonzero Chern numbers of the Weyl nodes, magnetic WSs are distinguished from ordinary ferromagnets by a lack of spin-dependent charge carrier scattering (extrinsic factor) and Fermi-surface contributions to the AHE. Instead, the AHE in ideal WSs (with two
Weyl nodes in the vicinity of the Fermi level) is purely intrinsic and determined only by the distance between the Weyl nodes in momentum space \cite{Burkov_AHE}. However, this is true only for the type-I WSs which have a point-like Fermi surface, while the AHE in the type-II WSs with tilted conical spectrum around the Weyl node is not universal and can change sign as a function of the parameters quantifying the tilt \cite{Zyuzin_AHE}. This universality can also be violated in the nodal-line WSs, such as $\textrm{Co}_3\textrm{S}_2\textrm{Sn}_2$, where the gapped nodal lines contribution to the AHE may be higher than the impact of the Weyl nodes themselves \cite{LiuFelser}. Nevertheless, an ideal WS, possessing purely topological AHE without nodal lines or magnetic moment contributions, can be found among spin gapless compensated ferrimagnets (e.g., $\mathrm{Ti}_{2}\mathrm{MnAl}$ \cite{Shi}). Notice that the cubic lattice symmetry of the typical magnetic WS crystals, such as pyrochlore iridates \cite{Wan,Sushkov_opt_exp}, enforces vanishing of the AHE due to the absence of a preferred axis. Nevertheless, the AHE can be recovered by applying a uniaxial strain that lowers the symmetry \cite{Yang_AHE}. 

The effects caused by WSs' nontrivial topology manifest in the optical \cite{VazifehFranz,Ashby-Carbotte,HosurQi:optics,Trivedi_Kerr,MaPesin,Tewari_CMEOpt,Tanaka_photo,Lee_photo, Minic_EM, ZyuzinENZ} and electron density responses \cite{Liu-CX,Burkov-pl,ZhouChangXiao,LvZhang,DasSarma_Tpl1,Kharzeev,DasSarma-Tpl2,ZhouChangXiao,Rosenstein,Ferreiros,DasSarma_SPP,Rudner_arcPl,Polini_arcPl,Sukhachov,Belyanin}. In particular, the AHE and CME give rise to gyrotropic terms in dielectric function \cite{Tatara_YAP}, which lead to the Faraday and Kerr magneto-optical effects in $\cT$-broken WS \cite{HosurQi:optics,Trivedi_Kerr} and to the natural
optical activity in $\cP$-broken WSs \cite{MaPesin,Souza_OptAct,Tewari_CMEOpt}. Moreover, the AHE, CME, and corresponding photocurrents in WSs can be generated by illuminating with circularly polarized light \cite{Tanaka_photo,Ran_photo,Lee_photo}. Nontrivial topology of $\cT$-broken WSs also results in the chiral Fermi arc plasmons with hyperbolic isofrequency contours \cite{Rudner_arcPl,Polini_arcPl}, in the chiral electromagnetic (EM) waves propagating at the vicinity of the magnetic domain wall in WSs \cite{Zyuzin}, in the transverse EM waves in a static magnetic field (helicons) \cite{Pellegrino}, and in the unusual EM modes with a linear dispersion in a neutral WS \cite{Rosenstein,Ferreiros}. Besides, the AHE  also makes the surface plasmon polaritons (SPPs) in WS chiral without applying an external magnetic filed (compare with Ref.~\cite{SongRudner_pl}). Particularly, in Ref.~\cite{DasSarma_SPP}, the behavior of SPP on the surface of WSs is calculated at different orientations of the AHE vector ($\textbf{b}$) and the direction of SPP propagation ($\textbf{q}$). The existence of the nonreciprocal SPP in WS, whose dispersion depends on the sign of the wave vector, is predicted in the Voigt configuration, when both $\textbf{b}$ and $\textbf{q}$ are in the plane of WS film, but perpendicular to each other.

The nonreciprocal unidirectional EM waves are widely known in magneto-optics, and dielectric waveguides (WGs) with ferrite cores or substrates, as well as films of magnetic dielectrics (MDs) (see Refs.~\cite{Tien_Rev1977, UFN1984, Camley_Rev1987}) are usually used for their transmission. Nonreciprocal optical elements are used in optical radiation control systems to create unidirectional optical circuits \cite{Popkov_Rev2005}, for the directed excitations in a ring laser \cite{Kravtsov}, in a laser gyroscope to eliminate the capture of the frequencies of counterpropagating modes \cite{Zvezdin_book}, as well as in fiber optic gyroscopes for the initial phase shift between the counter waves \cite{Petermann}. The theoretical description of nonreciprocal SPP was given in Refs.~\cite{ChiuQuinn},\cite{Wallis}, and the generalization for nonreciprocal WG modes in a film in the Voigt configuration was made in Ref.~\cite{Kushwaha}. Notice that in the Faraday configuration (magnetic field is along the propagation direction, parallel to the film) also the WG modes may exist but they will be reciprocal \cite{miyahara1988, ivanov1999}. For the design of compact optical elements with strong nonreciprocity effects, which do not need external sources of magnetic field, it is better to use materials with strong AHE and good WG properties. On the one hand, the MDs \cite{Belotelov} or delute magnetic semiconductors \cite{Rev_Dietl2014, fukumura2005} may be good WGs but they possess weak AHE. On the other hand, the ferromagnets may have strong AHE but they allow only anomalous light penetration, while ordinary dielectric response is suppressed due to the large electronic band gap. A compromise solution to this problem could become the Dirac (Weyl) ferromagnets, where, as it was shown in our previous work \cite{KotovBDS}, the weakly damped WG modes may arise due to the gapless spectrum.

In this paper, we propose to consider ferromagnetic WSs as the best candidates for the material which combines the WG properties of MDs or magnetic semiconductors with strong AHE in ferromagnets. Thanks to the strong dielectric response caused by the gapless Weyl spectrum and the large Berry curvature coming from the entangled Bloch electronic bands with SOC, ferromagnetic WSs may demonstrate good WG properties together with strong AHE, even stronger than in ordinary ferromagnets. We study the propagation of light in ferromagnetic WS films in the Voigt configuration without an external magnetic field. The role of a magnetic field plays the AHE in WS. We predict not only the nonreciprocity of the SPP on both sides of a WS film but also the existence of the nonreciprocal WG EM modes inside the film. The dispersions of the WG modes were obtained within the two-band model, accounting for the gapless nature of the Weyl spectrum. We also underline the key role played by the optical contrast between the media surrounding the film in the nonreciprocity magnitude of the predicted WG modes. Besides, the possibilities of varying of the nonreciprocity magnitude and operation frequencies of these modes by tuning the Fermi level in WS are discussed. Finally, we compare the AHE parameters in some real WS compounds. Our calculations show that WSs may become a good platform for the compact and tunable optical elements with strong nonreciprocity.

\section{Weyl semimetals optical response} \label{Sec2}
\begin{figure}[t]
	\centering
	\includegraphics[width=1\columnwidth]{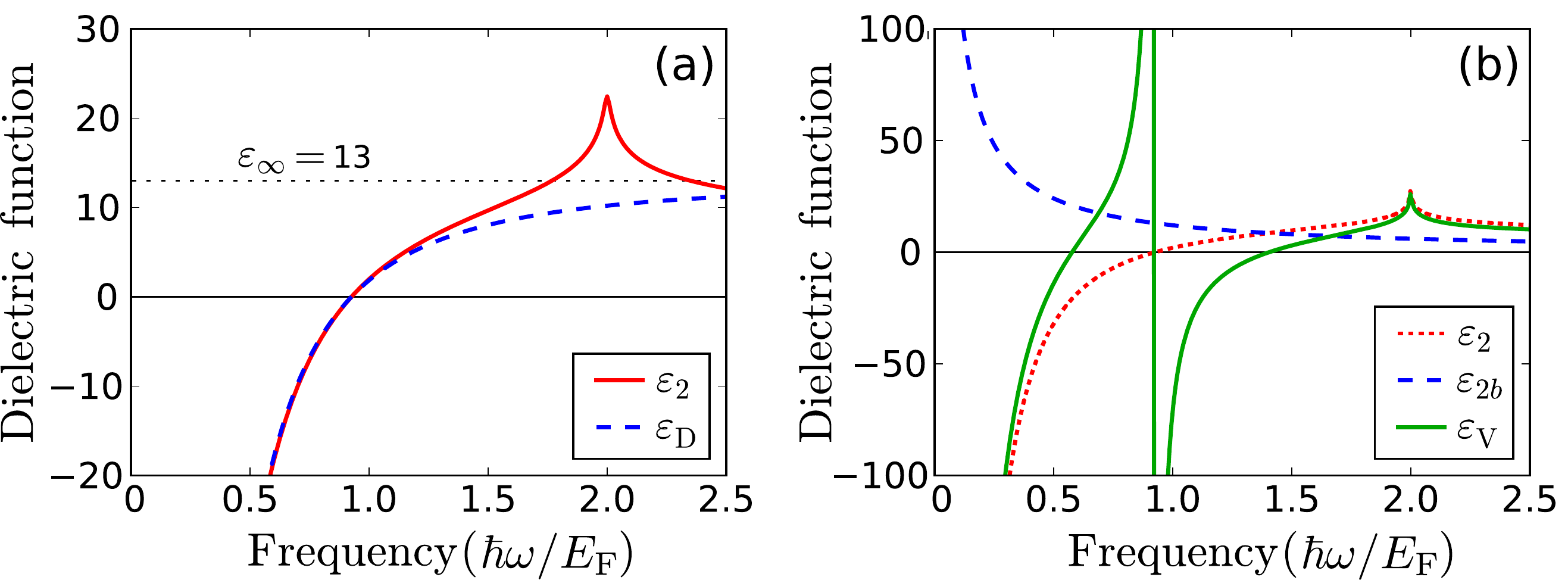} 
	\caption{\label{eps} (a) The dispersion of BDS dielectric functions in the one-band (Drude) $\varepsilon_\textrm{\tiny D}$ Eq.~(\ref{epsD}) and two-band $\varepsilon_2$ Eq.~(\ref{epsBDS}) models. (b) The dispersion of the components of WS dielectric tensor [diagonal component $\varepsilon_2$, nondiagonal component caused by the AHE $\varepsilon_{2b}$ Eq.~(\ref{eps2b})] and the Voigt dielectric function $\varepsilon_{\rm V}$. The parameters of WS  are set as $E_{\rm F}=150$meV, $v_{\rm F}=10^6$m/s, $g=24$, $\varepsilon_c=3$, $\varepsilon_b=6.2$, $\varepsilon_\infty=13$, $\Omega_b=\Omega_{\rm p}\approx0.93$ (i.e., $2b\approx0.4\rm\AA^{-1}$ and $\varepsilon_{2b}(\Omega=\Omega_{\rm p})=12$).     
	}
\end{figure}

Generally, the nonreciprocity effects in the Voigt configuration, as well as the magneto-optical effects, arise in a $\cT$-broken media, and the violation of $\cP$ symmetry leads to the natural optical activity effects, like in chiral media. In the case of BDS, the breaking of $\cT$ or $\cP$ symmetry splits each doubly degenerated Dirac point into a pair of Weyl nodes of opposite chirality, which are separated in the momentum space by vector $2\textbf{b}$ or in energy space by $2\hbar b_0$ (the chiral chemical potential). In the first case, there can be the AHE with the currents across the electric field, and in the second case the CME may occur with the currents induced along the magnetic field. The manifestation of WS topological nature in the optical response can be described by the additional axion term in the EM action \cite{Wilczek,ZyuzinBurkov,MacDonald}:
\begin{equation}
S_\theta=-e^2\big/\!\left(4\pi^2\hbar c\right)\int{dt\,d^3r\,\theta(\textbf{r},t)\,\textbf{E}\!\cdot\!\textbf{B}},
\end{equation}
where $\theta(\textbf{r},t)=2(\textbf{b}\!\cdot\!\textbf{r}-b_0t)$ is the axion angle. Varying this axion action with respect to the EM vector potential $\textbf{A}$ we
obtain the corresponding currents
\begin{equation}
\textbf{j}_\theta=\delta S_\theta/\delta\textbf{A}= -e^2\big/\!\left(4\pi^2\hbar\right)\left[\boldsymbol{\nabla}\theta(\textbf{r},t)\times\textbf{E}+\dot{\theta}(\textbf{r},t)/c\,\textbf{B}\right],
\end{equation}
where the first term corresponds to AHE  and the second one to the CME currents. These currents result in additional terms of the displacement vector \cite{DasSarma_SPP}:
\begin{equation}
\textbf{D}=\varepsilon\textbf{E}+\frac{{ie^2}}{{\pi\hbar\omega}}2\textbf{b}\times\textbf{E}-\frac{{ie^2}}{{\pi\hbar\omega c}}2b_0\textbf{B}. \label{D} 
\end{equation} 
Thus, to account for WS topological properties in the optical response, one may use the standard form of Maxwell equations with $\textbf{D}=\widehat\varepsilon_2\textbf{E}$ taking WS dielectric tensor in the form 
\begin{align}
\widehat\varepsilon_2=\begin{pmatrix}
\varepsilon_2 & 0 & 0  \\
0 & \varepsilon_2 & i\varepsilon_{2b}\\
0 & - i\varepsilon_{2b} & \varepsilon_2\\
\end{pmatrix}, \label{eps_tensor}
\end{align} 
where $\varepsilon_2$ is a BDS dielectric function and $\varepsilon_{2b}$ is a nondiagonal component caused by the AHE and CME. Since the nonreciprocal properties are always associated with the Hall response, we will consider the ferromagnetic WS in an equilibrium state without any external fields, with Weyl nodes separated only in momentum space (i.e., $b_0=0$). For this case, as follows from Eq.~(\ref{D}), the nondiagonal component of the tensor Eq.~(\ref{eps_tensor}) can be written as 
\begin{equation}
	i\varepsilon_{2b}=i2be^2\big/\!\left(\pi\hbar\omega\right)=i\epsilon_\infty\Omega_b\big/\Omega, \label{eps2b}
\end{equation}
where $\Omega_b=2br_{\rm s}\big/\!\left(k_{\rm F}\pi\varepsilon_\infty\right)$, $\Omega=\hbar\omega/E_{\rm F}$, $E_{\rm F}$ is the Fermi level, $k_{\rm F} =E_{\rm F}/\hbar v_{\rm F}$ is the Fermi momentum, $v_{\rm F}$ is the Fermi velocity, and $\varepsilon_\infty$ is the effective dielectric constant taking into
account all interband electronic transitions. In Ref.~\cite{DasSarma_SPP}, the standard one-band Drude model was used for $\varepsilon_2$, accounting only for the intraband electronic transitions:
\begin{equation}
\varepsilon_\textrm{\tiny D} =\varepsilon_\infty\left(1-\Omega_{\rm p}^2\big/\Omega^2\right), \label{epsD}
\end{equation} 
where $\Omega_{\rm p}^2=2r_{\rm s}g\big/\!\left(3\pi\varepsilon_\infty\right)$ denotes the bulk plasma frequency constant normalized to the Fermi level, $r_{\rm s} =e^2\!\big/\hbar v_{\rm F}$ is the effective fine structure constant, and $g$ is the degeneracy factor (the number of nondegenerated Weyl nodes). To describe the dielectric response in BDS more accurately, one should use the two-band model, taking into account the interband electronic transitions in the Dirac cone. As we have shown in Ref.~\cite{KotovBDS}, according to this model, a BDS dielectric function in the local response approximation at zero temperature has the form
\begin{equation}
\varepsilon_2=\varepsilon_b-\frac{2r_{\rm s}g}{3\pi}\frac{1}{\Omega^2} +\frac{r_{\rm s}g}{6\pi}\left[\ln\left(\frac{4\Lambda^2}{\left|\Omega^2-4\right|}\right)+i\pi\rm{\theta}(\Omega-2)\right], \label{epsBDS}
\end{equation}
where $\Lambda=E_{\rm c}/E_{\rm F}$ ($E_{\rm c}$ is the cutoff energy beyond which the Dirac spectrum is no longer linear), $\varepsilon_b$ is the effective background dielectric constant accounting the contributions from all bands below the Dirac cone. In Ref.~\cite{KotovBDS}, we obtained $\varepsilon_b=6.2$ for $g=24$ and $\epsilon_\infty=13$ ($\textrm{Eu}_2\textrm{IrO}_7$ \cite{Sushkov_opt_exp}). The difference between Eqs.~(\ref{epsD}) and (\ref{epsBDS}) [see Fig.~\ref{eps}(a)] is manifested at frequencies above the Fermi level when the dielectric behavior ($\varepsilon>1$) occurs and WG modes can exist. 

Notice that all the Weyl nodes have an equal contribution to the diagonal component of the dielectric tensor, which after summation gives the $g$-factor in Eq.~(\ref{epsD}). In contrast, to calculate the nondiagonal component, strictly speaking, one should integrate the Berry curvature over all occupied states in the first Brillouin zone, and Weyl nodes with different Chern numbers may even compensate each other, resulting in the vanishing of the AHE \cite{Yang_AHE}. Moreover, in the nodal-line WSs with strong AHE, such as $\textrm{Co}_3\textrm{S}_2\textrm{Sn}_2$, the gapped nodal lines contribution to the integrated Berry curvature is higher than the impact of the Weyl nodes themselves, and the anomalous Hall conductivity is more determined by the shape of the nodal lines than by the Weyl nodes separation \cite{LiuFelser}. Thus, strictly speaking, Eq.~(\ref{eps2b}) describes the contribution of only a one Weyl pair. To account for all the pairs and other possible sources of the Berry curvature the nondiagonal component given by Eq.~(\ref{eps2b}) should be multiplied by a coefficient depending on the Brillouin zone topology of a particular compound, which in general may be not directly expressed through the $g$-factor. For example, for $\textrm{Co}_3\textrm{S}_2\textrm{Sn}_2$ the numerical calculations and experimental measurements of the anomalous Hall conductivity gives the value about $1130\;\rm{S/cm}$ \cite{LiuFelser}, while the expression $\sigma_{\rm AH}=2be^2\big/\!\left(4\pi^2\hbar\right)$ following from  Eq.~(\ref{eps2b}) gives at $2b=0.47\AA^{-1}$ about $290\;\rm{S/cm}$, which is approximately four times lower. Nevertheless, Eq.~(\ref{eps2b}) can be used as a good estimation of the minimum value of the nonvanishing AHE.
			
\section{Nonreciprocal waves} \label{Sec3}
\begin{figure}[t]
	\centering
	\includegraphics[width=1\columnwidth]{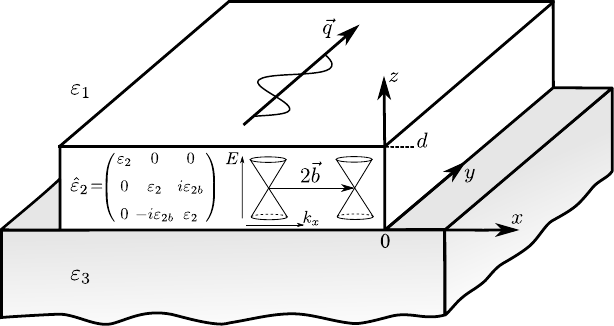} 
	\caption{\label{config}   
		The schematics of the nonreciprocal EM wave propagation in the WS film. In the Voigt configuration, the wave vector $\textbf{q}||\textbf{y}$ lies in the plane of the film but perpendicular to the AHE vector $2\textbf{b}||\textbf{x}$ separating the Weyl nodes in momentum space. $\widehat\varepsilon_2$ is the  dielectric tensor of WS, $\varepsilon_1$ is the free space dielectric constant, and $\varepsilon_3\neq\varepsilon_1$ is the dielectric function of a thick substrate. 
	}
\end{figure}

Let us consider the propagation of EM waves in the WS film with Weyl pairs where, in each pair, nodes are separated in momentum space by the wave vector $2\textbf{b}$. In the Voigt configuration, where nonreciprocal solutions can be found, the waves propagate in the plane of the film but perpendicular to the magnetic field (see Fig.~\ref{config}). In our case, the AHE plays the role of \textquotedblleft internal magnetic field\textquotedblright with the direction determined by the vector $\textbf{b}$. The WS film is asymmetrically bounded by two semi-infinite dielectric media with dielectric functions $\varepsilon_1$ and $\varepsilon_3$. As it will be shown below, for the nonreciprocal WG modes, it is important that $\varepsilon_1\neq\varepsilon_3$. The wave equation $\boldsymbol{\nabla}\times\left(\boldsymbol{\nabla}\times\textbf{E}\right)-k_0^2\widehat\varepsilon_2\textbf{E}=0$ with the vacuum wave vector $k_0=\omega/c$ for the considered system in the Voigt configuration has the form
\begin{align}
\begin{pmatrix}
q^2+k_{\rm V}^2-k_0^2\varepsilon_2 & 0 & 0  \\[0.5em]
0 & k_{\rm V}^2-k_0^2\varepsilon_2 & qk_{\rm V}-k_0^2i\varepsilon_{2b}\\[0.5em]
0 & qk_{\rm V}+k_0^2i\varepsilon_{2b} & q^2-k_0^2\varepsilon_2
\end{pmatrix}\begin{pmatrix} E_x \\[0.5em] E_y\\[0.5em] E_z\end{pmatrix}=0,
\label{Maks_V}
\end{align} 
where $\textbf{q}||\textbf{y}$ is the wave vector of EM waves, $k_{\rm V}=\sqrt{k_0^2\varepsilon_{\rm V}-q^2}$ and $\varepsilon_{\rm V}=\varepsilon_2-\varepsilon_{2b}^2\big/\varepsilon_2$ are the Voigt wave vector and dielectric function, respectively, which determine a light behavior inside the film in the considered configuration. This function has the resonance at the plasma frequency ($\varepsilon_2=0$), which leads to the splitting of the WG region ($\varepsilon_{\rm V}>1$) into two parts: the lower one is below the plasma frequency and the upper one is above it [Fig.~\ref{eps}(b)]. The lower region, where the anomalous light penetration into a metal at frequencies below the plasma one occurs, is typical for any magnetoplasma system in the Voigt configuration and is connected with the modification of the plasma frequency by the cyclotron resonance (see, e.g., Ref.~\cite{huang2008}). Interestingly, this phenomenon is accompanied by the effect of negative refraction, which can be observed not only in metamaterials but also in any gyrotropic (magnetic or chiral) system \cite{Tretyakov,Pendry,Mackay,Agranovich}. This effect can also be observed in WSs, which has been recently predicted in Ref.~\cite{Saito,Hayata}. The upper WG region is the manifestation of the dielectric response in any MD, magnetic semiconductor, or magnetic semimetal and may exist not only in the Voigt configuration \cite{miyahara1988, ivanov1999}. So, semiconductors or semimetals in an external magnetic field \cite{huang2008, miyahara1988, ivanov1999} or with intrinsic magnetic moment, such as dilute magnetic semiconductors \cite{Rev_Dietl2014}, in the Voigt configuration may possess both lower and upper WG regions. However, for the design of the WGs with strong nonreciprocity effects, which do not need external sources of magnetic field, it is better to use ferromagnets, where the AHE is much stronger than in MDs \cite{Belotelov} or magnetic semiconductors \cite{fukumura2005}. Thus, ferromagnetic WSs are the most suitable materials for these purposes. On the one hand, unlike ordinary ferromagnets, due to the gapless Weyl spectrum they possess a strong dielectric response and corresponding upper WG region, but on the other hand, unlike ordinary magnetic semiconductors, WSs due to the large Berry curvature have very strong AHE, even stronger than in ordinary ferromagnets \cite{LiuFelser}. Notice that in the Voigt configuration, the TE-polorized (s) EM waves will not feel the AHE, like in the case of external magnetic field, the carriers drifting parallel to the applied field do not experience a magnetic force. So, all the above comments are related to TM-polorized (p) WG modes.
 
Thus, we consider only the TM waves with field components $H_x$, $E_y$, $E_z$, and magnetic field in the form $H_x(r,t)=H_x(z)e^{i(qy-\omega t)}$, where $H_x(z)$ in the media with $\varepsilon_1$ ($z>d$), $\widehat\varepsilon_2$ ($0<z<d$), and $\varepsilon_3$ ($z<0$) (see Fig.~\ref{config}) is expressed as $H_{1x}(z)=H_1e^{-k_1z}$, $H_{2x}(z)=H_2e^{ik_{\rm V}z}+\widetilde{H_2}e^{-ik_{\rm V}z}$, and $H_{3x}(z)=H_3e^{k_3z}$, respectively. These fields correspond to the WG modes propagating inside the film with the Voigt wave vector $k_{\rm V}$ and exponentially decaying out of it. Employing the boundary conditions at the two interfaces $z=d$ and
$z=0$, we obtain the dispersion relation for the TM waves in the Voigt configuration (compare with Ref.~\cite{Kushwaha}):
\begin{align}
&\left[k_1 k_3\left(\varepsilon_2^2-\varepsilon_{2b}^2\right)
+k_2^2\varepsilon_1 \varepsilon_3 \pm q\varepsilon_{2b}\left(k_1\varepsilon_3-k_3\varepsilon_1\right)\right] \tan\left(k_{\rm V}{d}\right) \nonumber\\[0.3em]
&+k_{\rm V}\varepsilon_2\left(k_1 \varepsilon_3+k_3\varepsilon_1\right)=0,  \label{nonr}
\end{align} 
where $k_{1,2,3}=\sqrt{q^2-k_0^2\varepsilon_{1,2,3}}$ and $\pm$ sign before $q$ corresponds to the forward and backward propagation directions. Thus, the TM waves in the considered configuration will be nonreciprocal, which means that at certain frequencies they may propagate only forward ($\rm p_>$) but at another frequencies only backward ($\rm p_<$). For BDS films without Weyl features, the dispersion relations have a standard form \cite{KotovBDS}: for the TM waves,
\begin{align}
\left[{k_1 k_3\varepsilon_2^2-k_2^2\varepsilon_1\varepsilon_3} \right]\tan\left({k_2d}\right)+k_2\varepsilon_2\left({k_1\varepsilon_3+k_3\varepsilon_1}\right)=0,  \label{TM}
\end{align}
and for the TE waves, 
\begin{align}
\left[{k_1 k_3-k_2^2} \right]\tan\left({k_2d}\right)+k_2\left({k_1+k_3}\right)=0, \label{TE}
\end{align}
where $k_2=\sqrt{k_0^2\varepsilon_2-q^2}$ and $k_{1,3}=\sqrt{q^2-k_0^2\varepsilon_{1,3}}$. In WS film, the TE waves obey Eq.~(\ref{TE}) but the TM waves defined by Eq.~(\ref{nonr}), which in the absence of the AHE turns into Eq.~(\ref{TM}) in the limit $\varepsilon_{2b}\rightarrow0$ and by successive substitutions: $k_2\rightarrow ik_2$, then $k_{\rm V}\rightarrow k_2$. The dispersion of SPP in WS or BDS films can be obtained from Eqs.~(\ref{nonr}) or (\ref{TM}) by substitution $k_{\rm V}\rightarrow ik_{\rm V}$ and $k_2\rightarrow ik_2$, respectively.

\begin{figure}[t]
	\centering
	\includegraphics[width=1\columnwidth]{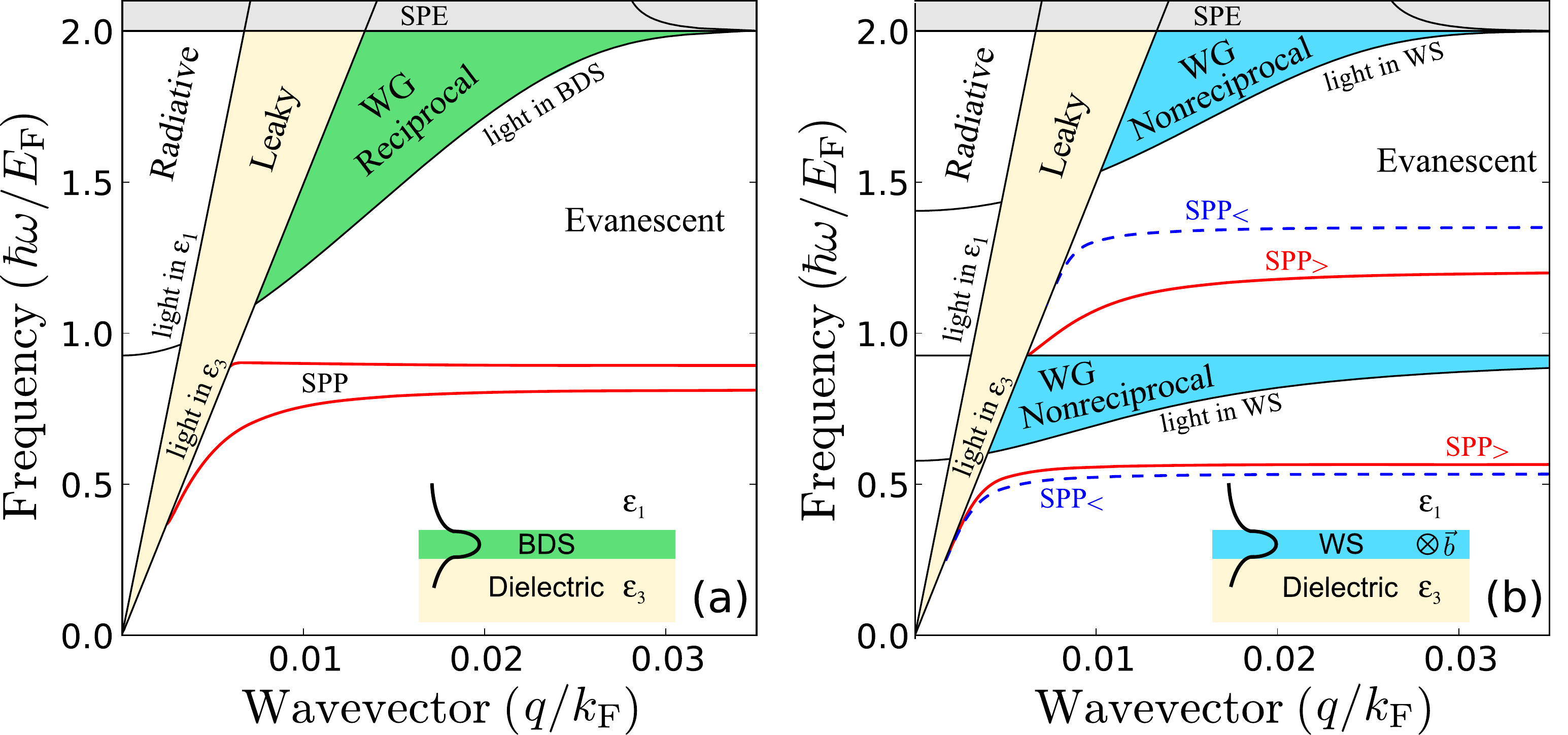} 
	\caption{\label{disp_shaded}   
		The dispersions of light and SPP in BDS ($\varepsilon_2$) film (a) and WS ($\widehat\varepsilon_2$) film in the Voigt configuration (b) with thicknesses $d=0.5\,\mu$m on the semi-infinite dielectric substrate with $\varepsilon_3=4$, while the medium above the films with $\varepsilon_1=1$. $\rm{SPP}_>$ and $\rm{SPP}_<$ are the forward and backward nonreciprocal SPP, respectively. SPE denotes the interband Landau damping region. The parameters of BDS and WS are the same as for Fig.~\ref{eps}.   
	}
\end{figure}

Using Eqs.~(\ref{eps_tensor}), (\ref{eps2b}), (\ref{epsBDS}), and (\ref{nonr})--(\ref{TE}) on Fig.~\ref{disp_shaded} at the same model parameters as for Fig.~\ref{eps}, we compared the dispersions of light and SPP in BDS ($\varepsilon_2$) film and WS ($\widehat\varepsilon_2$) film in the Voigt configuration with thicknesses $d=0.5\,\mu$m on the semi-infinite dielectric substrate ($\varepsilon_3=4$). For the case of BDS [Fig.~\ref{disp_shaded}(a)] we reproduced our previous result from Ref.~\cite{KotovBDS}, obtaining the WG region at $qc\big/\sqrt{\varepsilon_3}>\omega>qc\big/\sqrt{\varepsilon_2}$, leaky waves region at $\omega>qc\big/\sqrt{\varepsilon_3}$, and high (in-phase) and low (out-of-phase) SPP branches at $qc\big/\sqrt{\varepsilon_2}>\omega$. Both WG modes and SPP in this case are reciprocal. In the case of WS  [Fig.~\ref{disp_shaded}(b)] for TM waves, we get the splitting of the WG region into two parts, one of which lies below the plasma frequency. Both of these parts may contain the nonreciprocal TM WG modes. We also obtain the two pairs of the nonreciprocal SPP branches, which agrees with the results from Ref.~\cite{DasSarma_SPP}. Moreover, in the upper pair the nonreciprocity effect is much larger. Remarkably, from Eq.~(\ref{nonr}) it follows that the nonreciprocity effect of TM waves in the WS film is determined not only by the component $\varepsilon_{2b}$, but also by the term $\left(k_1\varepsilon_3-k_3\varepsilon_1\right)$. Therefore, the nonreciprocity effect grows with the optical contrast $\left|\varepsilon_1-\varepsilon_3\right|$ between the media above and below the WS film. This can be understood from the fact that nonreciprocal SPP excited on both sides of the film will compensate each other if the media from both sides are the same. In the measure of the optical contrast between these media, the nonreciprocity effect will appear in the collective SPP or WG TM modes propagating along the film. To demonstrate these phenomena, we considered the case of ordinary contrast, when the WS film lies on a dielectric substrate, and the case of high contrast, when the substrate is metallic. 

\begin{figure*}[t]
	\centering
	\includegraphics[scale=0.38]{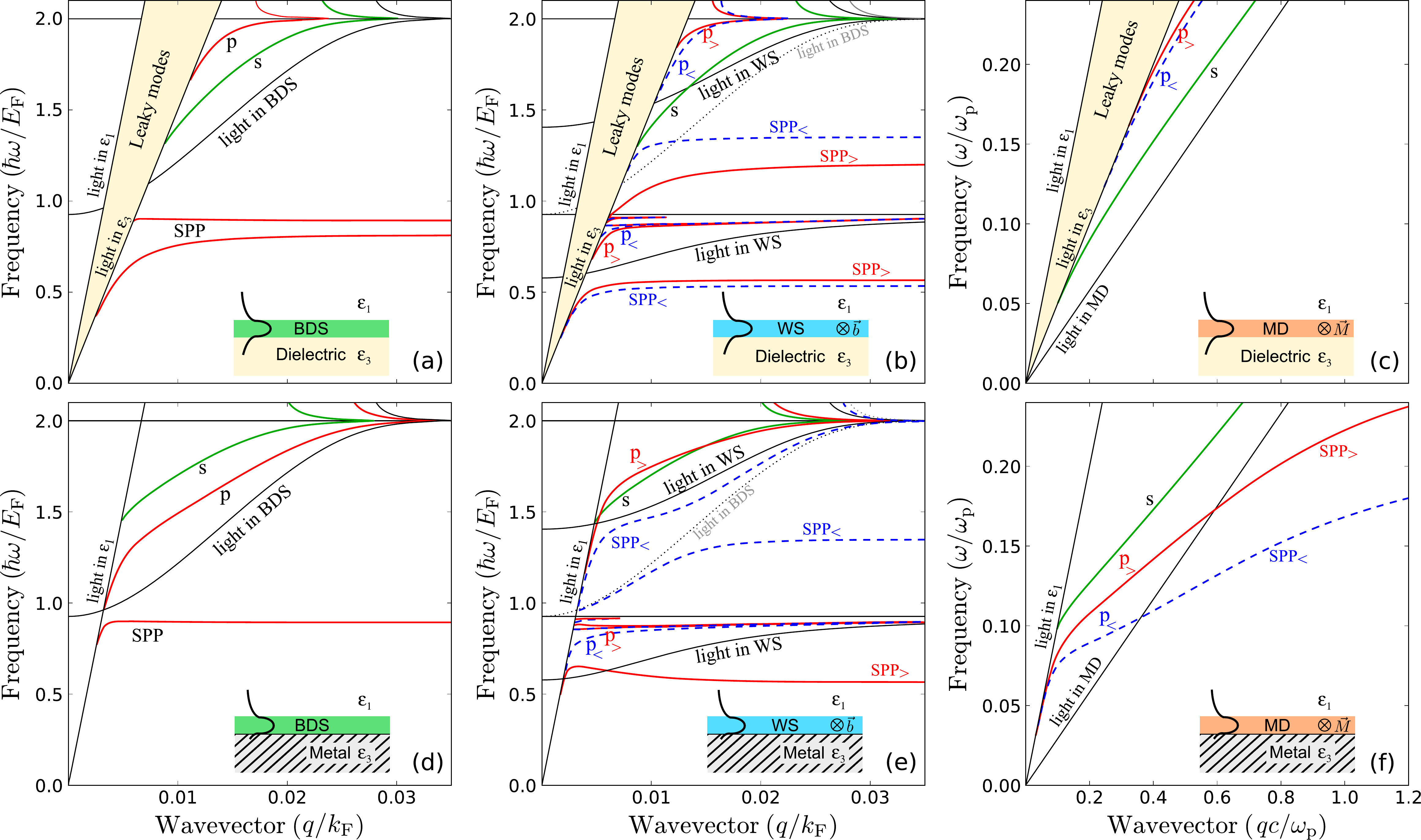} 
	\caption{\label{disp}   
		The dispersions of light and SPP in BDS ($\varepsilon_2$) film with $d=0.5\,\mu$m (a), in WS ($\widehat\varepsilon_2$) film with $d=0.5\,\mu$m and in MD film with $d=80\,$nm in the Voigt configuration. (a)-(c) the case of the ordinary contrast when all the films on the semi-infinite dielectric substrate ($\varepsilon_3=4$), (d)-(f) the case of the high contrast when all the films on the semi-infinite silver substrate ($\varepsilon_3=3.7-\Omega_{\rm p}^2\big/\Omega^2$, where $\Omega_{\rm p}=\hbar\omega_{\rm p}\big/E_{\rm F}$ and $\hbar\omega_{\rm p}=9.2eV$). The WG TE (s) modes do not feel the AHE and WG TM (p) modes in (b) and (c), (e) and (f) become nonreciprocal: ($\rm{SPP}_>$, $\rm{p}_>$) and ($\rm{SPP}_<$, $\rm{p}_<$) are the forward and backward nonreciprocal (SPP, WG TM modes), respectively. The MD dielectric tensor ($\widehat\varepsilon_2$) is the same as for the WS but with frequency independent components ($\varepsilon_\infty=13$, $\varepsilon_{2b}=4$). The medium above the films for all cases with $\varepsilon_1=1$. The parameters of BDS and WS are the same as for Fig.~\ref{eps}.   
	}
\end{figure*}
In the case of ordinary contrast, we compared the dispersions of WG modes and SPP in BDS and WS films with thickness $d=0.5\,\mu$m, as well as in the film of MD with thickness $d=80\,$nm and the same direction of magnetization as in WS. All the films are placed on a semi-infinite dielectric substrate ($\varepsilon_3=4$). The optical response of the MD we described by the same dielectric tensor Eq.~(\ref{eps_tensor}) as for the WS but with frequency independent components ($\varepsilon_\infty=13$, $\varepsilon_{2b}=4$). Comparing BDS and WS films [see Figs.~\ref{disp}(a) and \ref{disp}(b)], we get that the WG TM mode in WS becomes nonreciprocal and splits into two branches corresponding to the opposite directions of propagation, while the WG TE mode remains unchanged. There are also nonreciprocal TM waves in the WG region below the plasma frequency and two pairs of the nonreciprocal SPP in the evanescent region. In the MD film, certainly, there is no SPP and only one nonreciprocal WG region with a linear dispersion law [Fig.~\ref{disp}(c)].

In the case of metallic substrate (we take silver with $\varepsilon_3=3.7-\Omega_{\rm p}^2\big/\Omega^2$, where $\Omega_{\rm p}=\hbar\omega_{\rm p}\big/E_{\rm F}$ and $\hbar\omega_{\rm p}=9.2eV$ \cite{Shalaev}), at the considered frequencies its dielectric constant is very large and negative ($\varepsilon_3\sim-10^{3}$) which leads to a high optical contrast, and hence to a strong nonreciprocity effect. In the BDS film on the metallic substrate, the WG TM and TE modes swap places by frequency, and also only the high (in-phase) SPP branch exists [Fig.~\ref{disp}(d)]. In the WS film, on the metal there is really a large difference between the dispersion of the nonreciprocal waves propagating in the opposite directions [Fig.~\ref{disp}(e)]. In particular, the TM mode in one direction remains WG ($\rm p_>$) and in the opposite direction it becomes evanescent ($\rm{SPP}_<$). Also, in this case, the in-phase SPP splits to the pair of the nonreciprocal SPP with very high difference between $\rm{SPP}_>$ and $\rm{SPP}_<$. The similar behavior demonstrates the MD on the metal, where the WG TM and TE modes also swap places by frequency, the strong nonreciprocity effect of the WG TM modes takes place, and the nonreciprocal SPP arise due to the metallic substrate [Fig.~\ref{disp}(f)]. 
\begin{figure*}[t]
	\centering
	\includegraphics[scale=0.44]{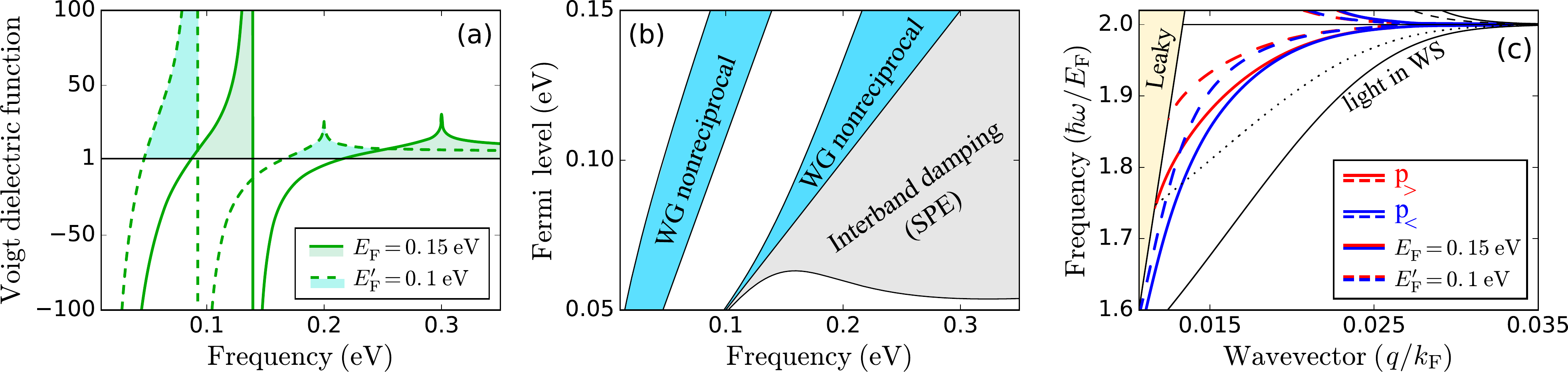} 
	\caption{\label{tuning} (a) The dispersion of the Voigt dielectric function $\varepsilon_{\rm V}$ in WS at different Fermi levels $E_{\rm F}=150$meV (solid line) and $E'_{\rm F}=100$meV (dashed line); the WG regions where $\varepsilon_{\rm V}>1$ are shaded by color. (b) The slice of the Voigt dielectric function vs. frequency and Fermi level at $\varepsilon_{\rm V}\left(\omega, E_{\rm F}\right)=1$; weakly damped WG modes regions are shaded by blue and the damping region is gray. (c) The dispersion of the nonreciprocal WG TM modes [forward $\rm{p}_>$ (red) and backward $\rm{p}_<$ (blue)] in WS film with $d=0.7\,\mu$m in the Voigt configuration at different Fermi levels $E_{\rm F}=150$meV (solid lines) and $E'_{\rm F}=100$meV (dashed lines); dotted line denotes the dispersion of light in WS at changed Fermi level $E'_{\rm F}$. Other parameters of WS are the same as for Fig.~\ref{eps}.     
	}
\end{figure*}

\section{Discussion} \label{Sec4}
For both ordinary and high optical contrasts between the media above and below the films, in the WS film the nonreciprocal WG modes and SPP can exist, similar to the waves which can be observed in WG with ferrite rods or MD films. However, in contrast to a MD film, in the WS film the WG mode frequency depends nonlinearly on the wave vector, and also there are two WG regions, one of which lies below the plasma frequency where the negative refraction in WS can be observed \cite{Saito}. But the main advantage of WS over MD films is the  magnitude of the nonreciprocity effect. In the WS, it depends not only on the surrounding media optical contrast, but also on the separation of the Weyl nodes in momentum space $2b=\Omega_bk_{\rm F}\pi\varepsilon_\infty/r_{\rm s}$. For all the figures, we took model parameters $E_{\rm F}=150$meV, $g=24$, $\Omega_b=\Omega_{\rm p}\approx0.93$, i.e., $2b\approx0.4\rm\AA^{-1}$ and $\varepsilon_{2b}(\Omega=\Omega_{\rm p})=12$. Such characteristics can be observed in the real compounds listed in Table~\ref{tab}. In WS, the separation of the Weyl nodes in momentum space can be so large $2b\approx0.5\rm\AA^{-1}$ ($\textrm{Co}_3\textrm{S}_2\textrm{Se}_2$) that the AHE dielectric tensor component $\varepsilon_{2b}\approx2.3/\hbar\omega[eV]$ even in the optical range ($\hbar\omega\sim2eV$) may be of the order of $\varepsilon_{2b}\sim1$. While for the typical MD film (bismuth iron garnet), in the optical range $\varepsilon_{2b}=0.003$ \cite{Belotelov} is by three orders of magnitude less than in some WS films. However, while bismuth iron garnet retains the magnetization at room temperature, all the WSs listed in Table~\ref{tab} can be used only at $T<T_{\rm C}\sim150K$. Nevertheless, WS such as compensated ferrimagnet   $\mathrm{Ti}_{2}\mathrm{MnAl}$ \cite{Shi}, with a
high Curie temperature $T_{\rm C}>650K$ and similar parameters as listed in Table~\ref{tab} for $\mathrm{Eu}_2\mathrm{IrO}_7$, may be the best candidate for room-temperature applications.

By tuning the Fermi level in WS, one can shift in frequency the nonreciprocal WG regions [see Fig.~\ref{tuning}(a)], but a number of limitations should be considered. First, the existence of the WG region assumes that $\varepsilon_{\rm V}=\varepsilon_2-\varepsilon_{2b}^2\big/\varepsilon_2>1$, i.e., at $\varepsilon_2>1$ for the upper region $\varepsilon_{2b}<\sqrt{\varepsilon_2^2-\varepsilon_2}\approx\left|\varepsilon_2\right|$ and at $\varepsilon_2<0$ for the lower region $\varepsilon_{2b}>\sqrt{\varepsilon_2^2-\varepsilon_2}\approx\left|\varepsilon_2\right|$. Thus, in the lower region, the AHE response dominates $\varepsilon_{2b}>\left|\varepsilon_2\right|$ and in the upper one the diagonal component should be larger $\varepsilon_{2b}<\left|\varepsilon_2\right|$. This leads to the different influence of the Fermi level tuning on the upper and lower WG regions. By making the slice of the Voigt dielectric function at $\varepsilon_{\rm V}\left(\omega, E_{\rm F}\right)=1$, we find that the upper WG region is much more sensitive to the Fermi level changes than the lower one [see Fig.~\ref{tuning}(b)]. In particular, the upper weakly damped WG modes region vanishes with the decrease of the Fermi level, while the lower one only slightly narrows. The damping region is located at $\hbar\omega>2E_{\rm F}$, where the interband Landau damping takes place [see imaginary part of Eq.~(\ref{epsBDS})]. Nevertheless, without lowering the Fermi level too low one can change the dispersion of the nonreciprocal TM WG modes in the upper region as well [see Fig.~\ref{tuning}(c)]. Besides, there is also an upper limit on the Fermi level: when it is high enough that the Fermi surfaces, enclosing the two Weyl nodes with opposite topological charges, merge, the magnitude of the AHE and corresponding WG modes nonreciprocity may dramatically change \cite{Burkov_AHE}. Thus, tuning the Fermi level in WS, one can vary the operation frequencies (which are near $\hbar\omega\sim E_{\rm F}$) of the predicted nonreciprocal waves in THz and mid-IR ranges. In particular, in WSs with a low Fermi level $E_{\rm F}\sim10$meV such as $\textrm{Eu}_2\textrm{IrO}_7$ \cite{Sushkov_opt_exp}, only the lower weakly damped nonreciprocal WG modes region can exist [see Fig.~\ref{tuning}(b)]; its frequencies lie in THz range, where $\varepsilon_{2b}\sim170$ (see Table~\ref{tab}) and the nonreciprocity can be very strong. However, as we discussed in Sec.~\ref{Sec3}, in contrast to the upper WG region, which is the manifestation of the dielectric response in WSs, the lower one lying below the plasma frequency may exist in any magnetoplasma system in the Voigt configuration. Nevertheless, in WSs with Fermi level around $E_{\rm F}\sim100$meV such as $\textrm{Co}_3\textrm{S}_2\textrm{Se}_2$ \cite{XuFelser}, both of the WG regions can exist and will belong to mid-IR frequency range, where the nonreciprocity is moderate ($\varepsilon_{2b}\sim10)$, as at the model parameters used for Figs.~\ref{eps}-\ref{tuning}. Notice that the nonlocal response in any materials may destroy the nonreciprocal effects \cite{Fan}, however, all our results for WSs were obtained at $q\ll k_{\rm F}$, where the local response approximation [see Eq.~(\ref{epsBDS})] works well.

\begin{table}[h]
\setlength\tabcolsep{0.5em}
	\begin{tabular}{l|ccccc}
		\hline\hline\toprule Compounds & $E_{\rm F}$(meV) & $g$ &  $2b$($\AA^{-1}$) & $\varepsilon_{2b}$ & $\omega_0$(THz) \\
		\hline\midrule
		\noindent
		\parbox[c]{1.8cm}{\raggedright\vspace{.2em}$\textrm{Y}_2\textrm{IrO}_7$ \cite{Wan}\\ $\textrm{Eu}_2\textrm{IrO}_7$ \cite{Sushkov_opt_exp}} & 10 & 24 & 0.37 & 170 & 2.4\\
		\hline\midrule
		$\textrm{Co}_3\textrm{S}_2\textrm{Sn}_2$ \cite{LiuFelser} & 60 & 12 & 0.47 & 36 & 14.5\\
		\hline\midrule
		$\textrm{Co}_3\textrm{S}_2\textrm{Se}_2$ \cite{XuFelser} & 110 & 12 & 0.5 & 21 & 26.6\\
		\hline\hline\bottomrule 
	\end{tabular}  
	\caption{\label{tab}
		The list of magnetic WSs with different Fermi levels $E_{\rm F}$, numbers of nondegenerated Weyl nodes $g$, separations of the Weyl nodes in momentum space $2b$, corresponding dielectric tensor AHE components $\varepsilon_{2b}$ taken at $\hbar\omega_0=E_{\rm F}$, and operation frequencies $\omega_0$.}
\end{table}

\section{Conclusion} \label{Sec5}

In summary, we predict the existence of nonreciprocal WG modes in ferromagnetic WS films in the Voigt configuration without an external magnetic field. The role of a magnetic field plays the AHE in WS, which, being purely intrinsic and universal in ideal WSs, depends only on the separation of the Weyl nodes in momentum space. The nonreciprocity value also depends on the optical contrast between the media surrounding a WS film, particularly, a metallic substrate leads to a significant increase of the nonreciprocity due to the high optical contrast with the medium above the film. We show that the nonreciprocal WG modes may exist in the two frequency regions: the lower one is below the WS plasma frequency and the upper one is above it. The lower WG region, where the negative refraction can be observed, is typical for any magnetoplasma system in the Voigt configuration, while the upper one is the manifestation of the dielectric response in WSs. We provide the AHE parameters of the real WS materials where a strong nonreciprocity can be observed even without a help of the surrounding media optical contrast. Such high values of the AHE in ferromagnetic WSs may be useful not only for the nonreciprocity but also for the gyrotropic effects. Moreover, tuning the Fermi level in WSs, one can vary the operation frequencies of the WG modes in THz and mid-IR ranges. We find that the upper WG region is much more sensitive to the Fermi level changes than the lower one. In particular, the upper weakly damped WG modes region vanishes with the decrease of the Fermi level, while the lower one only slightly narrows. So, to work with both WG regions, one should use WSs with rather high Fermi levels. Thanks to the strong dielectric response caused by the gapless Weyl spectrum and the large Berry curvature coming from the entangled Bloch electronic bands with SOC, ferromagnetic WSs combine the best WG properties of MDs or magnetic semiconductors with strong AHE in ferromagnets. Thus, ferromagnetic WSs allow one to realize giant tunable gyrotropic and nonreciprocity effects for a propagating light, which paves the way to the design of compact, tunable, and effective nonreciprocal optical elements. 

\section*{Acknowledgments}
The authors are grateful to \mbox{A. A. Sokolik} for useful discussions. The work was supported by the Russian Foundation for Basic Research (17-02-01322, 17-02-01134, 18-52-00002). Yu. E. L. thanks the Basic Research Program of the National Research University Higher School of Economics. 

%\bibliographystyle{apsrev4-1}%
%\bibliography{bibWS}%

%

\end{document}